\documentclass[
]{aa}

\usepackage[varg]{txfonts}
\usepackage{natbib}
\bibpunct{(}{)}{;}{a}{}{,} 
\usepackage[colorlinks=true,allcolors=blue]{hyperref}

\usepackage{physics}
\usepackage{siunitx}
\usepackage{physunits}
\DeclareSIUnit\gauss{G}
\DeclareSIUnit\erg{erg}

\usepackage{amsmath}
\usepackage{lipsum}
\usepackage{graphicx} 
\graphicspath{{./}{fig/}}
\usepackage{subfig}

\newcommand{\Ttot}[0]{T_\mathrm{Tot}}
\newcommand{\Tinf}[0]{T_\infty}
\newcommand{\Tmax}[1]{\max[T#1]}
\newcommand{\Ttotmax}[1]{\max[\Ttot#1]}
\newcommand{\pert}[1]{^{(#1)}}

\newcommand{\phiinf}[0]{\phi_\infty}
\newcommand{\phitot}[0]{\phi_\mathrm{Tot}}

\newcommand{\ebysus}{\texttt{Ebysus}}
\newcommand{\bifrost}{\texttt{Bifrost}}
\newcommand{\kappaeff}{\kappa_\parallel^\ast}

\newcommand{\change}[1]{{#1}} 

\newcommand{\mysz}{0.98}

\begin{document}

    
    \title{On Thermal Conduction in the Solar Atmosphere: 
    	       An Analytical Solution for \change{Nonlinear} Diffusivity without Compact Support}
    \titlerunning{On Thermal Conduction in the Solar Atmosphere}
    
    \author{
        S.V. Furuseth\thanks{\email{s.v.furuseth@gmail.com}}\inst{1,2}%
        \and
        G. Cherry\inst{1,2}%
        \and
        J. Mart\'inez-Sykora\inst{1,2,3,4}
        }

   \institute{
        Rosseland Centre for Solar Physics, University of Oslo, PO Box 1029 Blindern, 0315 Oslo, Norway
        \and 
        Institute of Theoretical Astrophysics, University of Oslo, PO Box 1029 Blindern, 0315 Oslo, Norway
        \and
        Bay Area Environmental Research Institute, Moffett Field, CA 94035, USA
        \and
        Lockheed Martin Solar and Astrophysics Laboratory, Palo Alto, CA 94304, USA
        }

   \date{
   	}

\abstract
   {
    The scientific community employs complicated \change{multi}physics simulations to understand the physics in Solar, Stellar, and Interstellar media. These must be tested against known solutions to ensure their validity. 
    Several well-known tests exist, such as the Sod shock tube test. However, a test for \change{non}linear diffusivity is missing.
    This problem is highly relevant in the Solar atmosphere, where various events release energy that subsequently diffuses by Spitzer thermal conductivity.
   }
   {
    The \change{aim is to derive} an analytical solution for \change{non}linear diffusivity in 1D, 2D, and 3D, which allows for a \change{nonzero} background value. 
    The solution will be used to design a test for numerical solvers and study Spitzer conductivity in the Solar atmosphere. 
   }
  {
    There existed an ideal solution assuming zero background value. We perform an analytical first-order perturbation of this solution.
    The first-order solution is first tested against a dedicated \change{non}linear diffusion solver, whereupon it is used to benchmark the single- and \change{multi}fluid radiative magnetohydrodynamics code \ebysus{}, used to study the Sun. 
    The theory and numerical modeling are used to investigate the role of Spitzer conductivity in the transport of energy released in a nanoflare.
   }
   {
    The derived analytical solution models \change{non}linear diffusivity accurately within its region of validity and approximately beyond. 
    Various numerical schemes implemented in the \ebysus{} code have been found to model Spitzer conductivity correctly.
    The energy from a representative nanoflare has been found to diffuse 9~Mm within the first second of its lifetime due to Spitzer conductivity alone, strongly dependent on the electron density.
   }
  {
	The analytical first-order solution is a step forward in ensuring the physical validity of intricate simulations of the Sun.
	Additionally, since the derivation and argumentation are general, they can easily be followed to treat other \change{non}linear diffusion problems. 
  }


   \maketitle
%



\nolinenumbers


\section{Introduction}
\label{sec_intro}
The scientific community of today relies heavily on complicated \change{multi}physics simulations. To increase the trustworthiness of such simulations, every single physics module should be benchmarked against analytical solutions. Several such benchmarks exist and are often used, such as the Sod shock tube test for hydrodynamics codes~\citep{sod1978}.

In this paper, we are interested in partial differential equations of parabolic terms (\change{non}linear diffusion) in Cauchy problems. Assuming radial symmetry, that is given by
\begin{equation}
    \pdv{T}{t} = \frac{1}{r^{s-1}}\pdv{}{r}
        \left(r^{s-1}{D(T)\pdv{T}{r}}\right),
    \quad D(T) = KT^n,
    \label{eq_dTdt_intro}
\end{equation}
where $K$ is a constant, ${n>0}$ is the (positive) \change{non}linearity exponent, and 
${s\in\{1,2,3\}}$ is the number of dimensions.
$T$ is chosen because it will later be temperature, but it can represent any value.
A diverse set of problems can be modeled by such \change{non}linear diffusion with different exponents $n$, a \change{in}exhaustive list is given by~\citet{diez1992}.

It is computationally demanding to solve such problems numerically. That is because the stability condition of an explicit diffusion solver requires the time step to scale like the spatial resolution squared, ${\Delta t\propto \Delta r^2}$ \change{\citep[see][p. 1044]{numrec2007}}.
Therefore, different algorithms have been developed to solve such problems and bypass this time-step constraint. These algorithms need to be tested, preferably against an analytical solution. 

An analytical solution exists for \change{non}linear diffusion of an instantaneous point source with zero background $T$, making the diffusion coefficient in Eq.~\eqref{eq_dTdt_intro} equal to zero beyond the extent of the point source~\citep{Pattle1959}. That derivation finds self-similar solutions that keep their shape with a gradually lower peak value and broader spatial extent with time.
In some problems, however, it is not realistic to have zero background value. In this paper, we extend this theory by a perturbation to include a \change{nonzero} background value. The theory will be used to analyze the efficacy of different numerical schemes.

\subsection{Thermal conductivity in the Solar atmosphere}
\label{sec_intro_TD}

We will apply the derivation to the modeling of thermal conductivity by electrons in a plasma, as in the Solar atmosphere~\citep{spitzer1962}. 
If we assume a negligible heat conduction perpendicular to the magnetic field in the plasma and a constant mass density $\rho$, 
the conductive term can be written on the form (see App.~\ref{app_thermalconductivity} for details)
\begin{equation}
    \left(\pdv{T}{t}\right)_\mathrm{cond} 
    = 
    \grad_\parallel\cdot \left( 
    \frac{\kappa_{\parallel}^{\ast}}{c_v \rho} 
    T^{5/2} \grad_\parallel T \right) ,
    \label{eq_dTdt_cond_intro}
\end{equation}
where $c_v$ is the specific heat capacity per mass, $\rho$ is the mass density, and ${\kappaeff T^{5/2}}$ is the parallel thermal conductivity with exponent ${n=\tfrac{5}{2}}$. Since the conduction is along the field lines, this is modeled as diffusion in ${s=1}$~dimensions.
The coefficient in the thermal conductivity is ${\kappaeff\sim 10^{-6}~\si{\erg\per\second\per\centi\meter\per\kelvin\tothe{7/2}}}$ for a fully ionized hydrogen gas in the solar atmosphere~\citep{spitzer1962,priest1984}.
To bypass the time step constraint of typical explicit methods, this term can be solved implicitly as in \bifrost{}~\citep{gudiksen2011}, with the wave method first implemented in \texttt{MURaM}~\citep{rempel2016}, or the explicit orthogonal Chebyshev method known as ROCK2~\citep{abdulle_rock2_2001,Abdulle2002,abdulle_2008,Zbinden2011} and implemented in \ebysus{}~\citep{sykora_2020}. 


One cannot reasonably assume a zero background temperature when modeling the \change{non}linear thermal conductivity in the Solar atmosphere. However, the background temperature can be much smaller than the source temperature, \change{for example} when modeling thermal conduction from the $\order{\si{\mega\kelvin}}$ corona to the $\order{\si{\kilo\kelvin}}$ photosphere or when modeling the localized release of energy from a nanoflare, as described and studied numerically by~\citet{testa2014,polito2018,bakke2022}.
\citet{polito2018} found that lower-energy electrons tend to release more energy in the corona than higher-energy electrons, as previously found in~\citet{Reep_2015}, but that thermal conduction is more effective at heating the magnetic loop. 
Further, they found that the initial conditions \change{(IC)} of the loops prior to a nanoflare, \change{in particular} temperature and density, significantly impact the atmosphere's response. 
This work has been extended into stellar flare events, where thermal conduction is more important than radiation and a key process in the energy flux~\citep{kowalski_stellarflare_2024}.
These studies modeled several physical processes that play a role in the energy flux in flares to get a realistic description, including electron beams, thermal conductivity, \change{non}local thermal equilibrium, and radiation. Each of these processes competes and it is crucial to understand their independent solutions to understand and separate them. 
We will focus on the role of thermal conductivity.

\subsection{Outline}
In this paper, we will first make an analytical solution of the \change{non}linear diffusivity in Sec.~\ref{sec_theory}, which we will verify numerically in Sec.~\ref{sec_testtheory}.
Based on this, we will explain and show how to use this derivation to benchmark a code in Sec.~\ref{sec_benchmark}.
Then, we will make a numerical experiment of how fast energy diffuses from a nanoflare in Sec.~\ref{sec_flares} before we conclude.

\section{Analytical derivation}
\label{sec_theory}


\subsection{Solution with zero background\texorpdfstring{, $\Tinf=0$}{, Tinf=0}}
\label{sec_theory_ideal}

\begin{figure}[t]
    \centering
    \includegraphics[width=0.95\columnwidth]{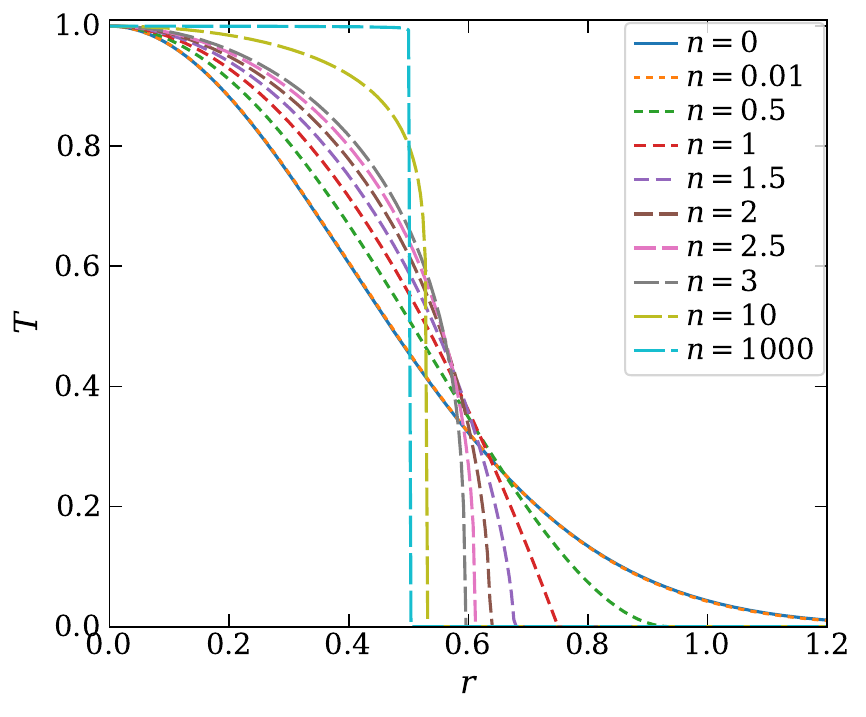}
    \caption{Illustration of self-similar solutions in ${s=1}$ dimensions for different ${n}$ given by the legend. All solutions have peak value ${T_0=1}$ and total quantity ${\phi_0=1}$ (note that the solution is symmetric around ${r=0}$).}
    \label{fig_selfsim_n}
\end{figure}

An initial instantaneous point source quantity $\phi_0$ released at ${t=0}$ and centered at ${r_0=0}$ will diffuse as the self-similar solutions first described by \citet{Pattle1959} and comprehensively derived in App.~\ref{app_selfsim}. If we allow for a finite initial extent, they can be written as
\begin{align}
    T(r,t) &= 
    \begin{cases}
        T_0\left(1+\chi t\right)^{-\tfrac{s}{s\,n+2}}
        \left(1-\dfrac{r^2}{R(t)^2}\right)^{\tfrac{1}{n}} 
        ,  & \text{if $r<R(t)$} \\ 
        0~, & \text{otherwise}
    \end{cases} \label{eq_T}
    \\
    R(t) &= R_0\left(1+\chi t\right)^{\tfrac{1}{s \, n+2}}, \label{eq_R} \\
    \chi &= \frac{s \, n+2}{n} \frac{2KT_0^n}{R_0^2} =  \frac{s \, n+2}{n} \frac{2D(T_0)}{R_0^2}, \label{eq_chi}
\end{align}
where ${T_0=\Tmax{(t=0)}}$ is the initial representative peak value
and $R_0$ is the initial representative width beyond which ${T=0}$.
Self-similar shapes for various $n$ in ${s=1}$ dimensions are displayed in Fig.~\ref{fig_selfsim_n}.
The initial peak and width are related to the total quantity (area under the graph) by
\begin{equation}
    \phi_0 = \int\limits^\infty T(r,0)\dd r^s 
    = \frac{\Omega_s B\left(\frac{s}{2},{\frac{1}{n}+1}\right)}{2}T_0 R_0^s
    \equiv \frac{T_0 R_0^s}{G_{s,n}}, \label{eq_phiG}
\end{equation}
where $\Omega_s=\{2,2\pi,4\pi\}$ is the solid angle for $s=\{1,2,3\}$ dimensions, 
${B\left(\tfrac{s}{2},{\tfrac{1}{n}+1}\right)}$
is Euler's Beta integral~\change{\citep[see][ch. 6]{abramowitz_1965}},
and ${G_{s,n}\lesssim 1}$ (for ${n\geq1}$) is a time-independent geometrical factor depending on $n$ and $s$. 

The representative peak $T_0$ and width $R_0$ are not clearly defined for an instantaneous point source. However, that is not a severe problem as any distribution that is 0 beyond a finite radius eventually will approach the shape in Eq.~\eqref{eq_T}. This can be understood by taking the limit ${\chi t \gg 1}$, corresponding to ${R_0\rightarrow 0}$, and using Eq.~\eqref{eq_phiG}, to get
\begin{align}
    \lim\limits_{\chi t \gg 1} T(r<R(t),t) 
    &= 
        \left(\dfrac{n}{s \, n+2}\dfrac{(G_{s,n}\phi_0)^{\tfrac{2}{s}}}{2Kt} \right)^{\tfrac{s}{s \, n+2}}
        \!\! \left(1-\dfrac{r^2}{R^2(t)} \right)^{\tfrac{1}{n}} ,
    \label{eq_T_chit}
    \\
    \lim\limits_{\chi t \gg 1} R(t) 
    &= \left( \frac{s \, n+2}{n}2K(G_{s,n}\phi_0)^n t\right)^{\tfrac{1}{s \, n+2}}
    . \label{eq_R_chit}
\end{align}
The distribution depends only on the initial total quantity $\phi_0$, not $T_0$ and $R_0$. The peak and width have simple dependencies on time, ${\Tmax{} \propto t^{-\frac{s}{s \, n+2}}}$ and ${R\propto t^{\frac{1}{s \, n+2}}}$, making ${\phi(t)=\Tmax{(t)}R(t)^s/G_{s,n}}$ a constant.

For comparison, in the limiting case ${n=0}$, corresponding to isotropic diffusion (${D=K}$), an initial instantaneous point source at ${r_0=0}$ is known to diffuse as a Gaussian that can be written as
\begin{align}
    T(r,t) 
    &= T_0 \left( 1 + \chi t \right)^{-\frac{s}{2}}
    \exp(-\frac{r^2}{2R_\sigma^2(t)}) , \label{eq_T_n0}\\ 
    R_\sigma(t) 
    &= R_{\sigma 0}\left(1 + \chi t\right)^{\tfrac{1}{2}}, \label{eq_R_n0}\\
    \chi &= \frac{2K}{R_{\sigma 0}^2} = \frac{2D(T_0)}{R_{\sigma 0}^2} .
    \label{eq_chi_n0}
\end{align}
Here, $R_\sigma(t)$ is the standard deviation of the Gaussian distribution, not the boundary beyond which ${T=0}$, since the Gaussian extends to infinity. The total quantity is
\begin{equation}
    \phi_0 = (2\pi)^{\frac{s}{2}}T_0 R_{\sigma 0}^s.
\end{equation}

We can see by the self-similar shapes in Fig.~\ref{fig_selfsim_n} that the Gaussian distribution is the limiting case for $n\rightarrow0^{+}$. As ${n\rightarrow\infty}$, the edges become sharper as the diffusion coefficient in Eq.~\eqref{eq_dTdt_intro} is relatively stronger at larger values, making the geometrical factor ${G_{1,n}\rightarrow 1/\Omega_{1}=0.5}$ in ${s=1}$ dimensions.

\subsection{Including a \change{nonzero} background, \texorpdfstring{$\Tinf>0$}{Tinf>0}}
\label{sec_theory_Tinf}

\begin{figure}
    \centering
    \includegraphics[width=.95\columnwidth]{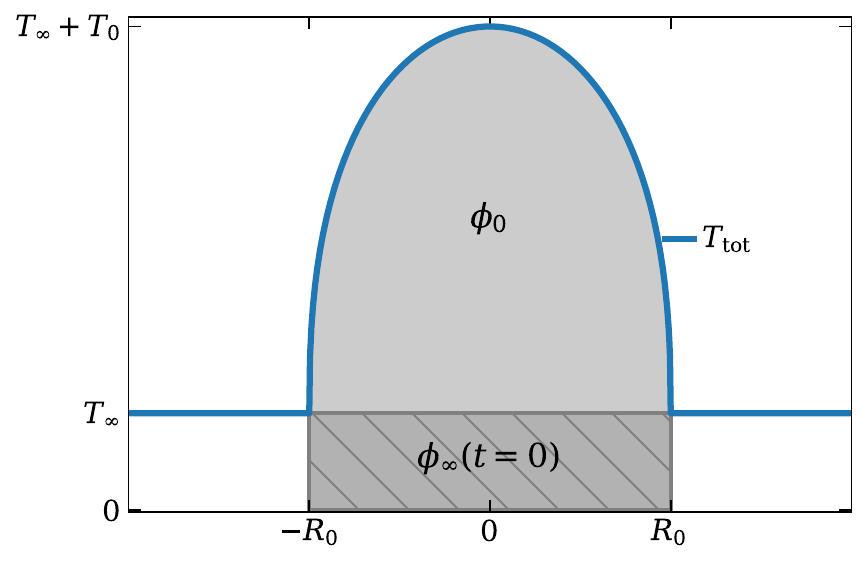}
    \caption{Illustration of \change{IC}s with a background value ${\Tinf\geq0}$ for ${s=1}$ dimensions and ${n=5/2}$.}
    \label{fig_Tinf}
\end{figure}

The aim of this paper is to include a background value ${\Tinf}$ and then predict how it evolves. By assuming that the background has had time to reach equilibrium, we set $\Tinf$ to a constant. That is, however, not essential for the following arguments.
The total initial distribution is given by
\begin{equation}
    \Ttot(r,0) = \Tinf + T(r,0),
    \label{eq_Ttot_0}
\end{equation}
as illustrated in Fig.~\ref{fig_Tinf}.
If one completely neglects the impact of the background value and assumes that the source evolves as before, one gets the ideal zeroth-order solution 
\begin{equation}
	\Ttot\pert{0}(r,t)
	= \Tinf + T(r,t; R_0, T_0),
	\label{eq_T_pert0}
\end{equation}
where the superscripted $(0)$ marks the zeroth order. 
This is a too na\"{i}ve approach. 
An important assumption in the ideal derivation in App.~\ref{app_selfsim}~\citep[based on][]{Pattle1959} was that $T$ eventually would go to zero beyond a finite radius, given by Eq.~\eqref{eq_ass_Tinf0}. That assumption is broken by including a background value ${\Tinf>0}$.
By understanding the derivation in App.~\ref{app_selfsim}, we realize in the following the importance of this constraint and the consequences of breaking it.

The shape of $T(r,t)$ given by Eq.~\eqref{eq_T} for ${r<R(t)}$ solves the nonlinear diffusion equation in Eq.~\eqref{eq_dTdt_intro}. Alternative solutions include the trivial solution ${T=0}$ and constant solution ${T=\mathrm{const}>0}$. The trivial solution has the important added benefit that it makes the diffusion coefficient in Eq.~\eqref{eq_dTdt_intro} zero, perfectly separating the two regions at the boundary ${r=R(t)}$. 
A \change{nonzero} background distribution ${\Tinf>0}$ will cause diffusion out beyond $R(t)$, preventing the perfect separation of the two regions. 

Consider first the case ${\Tinf\gg T_0}$. This allows a Taylor expansion of the diffusion coefficient around the background as
\begin{equation}
\begin{split}
    D(\Ttot) 
    &= K(\Tinf + T)^n \\
    &= K\Tinf^n\left[1 + n\frac{T}{\Tinf} + \frac{n(n-1)}{2}\frac{T^2}{\Tinf^2} + \order{\frac{T^3}{\Tinf^3}}\right].
\end{split}
\label{eq_D_taylor}
\end{equation}
This diffusion will be dominated by the first term, which is constant. The second term will gradually become less important as $\Ttotmax{(t)}$ will decrease \change{toward} $\Tinf$. Hence, the source will immediately diffuse into the Gaussian shape in Eq.~\eqref{eq_T_n0}.

Consider next the case ${\Tinf\ll T_0}$, which is less trivial and more interesting for our use case. It is tempting to make a similar expansion as in Eq.~\eqref{eq_D_taylor}, with $T$ and $\Tinf$ exchanged, giving terms of $T^n$, $T^{n-1}$, etc. That expansion is valid at ${r\ll R}$, where ${T\gg\Tinf}$. The \change{nonzero} background value causes a slightly larger diffusion coefficient, making the peak value $\Ttotmax{}$ drop slightly faster. 
However, at ${r\rightarrow\infty}$, the instantaneous point source has not yet diffused, making ${T=0\ll\Tinf}$ and the expansion singular. 
In the intermediate region, at ${r\sim R}$, the background and source become comparable, ${T\sim\Tinf}$, making the terms of different order in $T$ comparable as well. 
Thus, the distribution will not have as sharp edges as for ${n > 1}$ in Fig.~\ref{fig_selfsim_n}, it will have wider tails as for smaller values of ${n\rightarrow0}$.
Nevertheless, by considering the inverse time scale $\chi$ in Eq.~\eqref{eq_chi}, which is proportional to the peak diffusion coefficient, one can find that the global widening will still be dominated by the $T^n$ term. 
Hence, the overall widening will follow the $n$-solution, but it will be reshaped to have wider tails.
Nevertheless, as ${\chi t\rightarrow\infty}$, the peak value will eventually drop to ${\Tmax{(t)}\ll\Tinf}$, and it will diffuse into a Gaussian, as for the previous case, ${\Tinf\gg T_0}$.

\subsection{First-order solution for \texorpdfstring{$\Tinf\ll T_0$}{Tinf<<T0}}
\label{sec_theory_pert}

In the previous section, we discussed qualitatively how the solution will change when adding a \change{nonzero} background value ${\Tinf>0}$. 
In this section, we derive a first-order perturbation solution that quantitatively estimates it.
%
This is done assuming an identical shape to the ideal solution but with a modified peak and width. We consider separately the two time regimes, ${\chi t \ll1}$ and ${\chi t \gg 1}$. In the former, the zeroth-order solution depends on the shape of the \change{IC}, while in the latter, only on the source quantity $\phi_0$.

Consider first ${\chi t \ll1}$, the beginning of the diffusion. In this regime, we focus on how the peak value $\Ttotmax{}$ evolves. The peak of the zeroth-order solution in Eq.~\eqref{eq_T_pert0} is given by
\begin{equation}
    \Ttotmax{\pert{0}{(t)}} 
    = \Tinf + T(0,t) = \Tinf + T_0\left[1+\chi(T_0) t\right]^{-\tfrac{s}{s \, n+2}},
\end{equation}
where the dependence on $T_0$ in $\chi$ from Eq.~\eqref{eq_chi} has been made explicit. 
The first-order perturbation is to include the background value in $\chi$ as
\begin{equation}
\begin{split}
    \Ttotmax{\pert{1}{(t)}} 
    &= \Tinf + T_0\left[1+\chi(T_0+\Tinf) t\right]^{-\tfrac{s}{s \, n+2}} \\
    &= \Tinf + T_0\left[1+\left(1+\frac{\Tinf}{T_0}\right)^{n} 
    \change{\chi(T_0) t}
    \right]^{-\tfrac{s}{s \, n+2}}.
\end{split}
\end{equation}
We expand the parentheses by taking the limits ${\chi t\ll1}$ and ${\Tinf\ll T_0}$ to get a simple algebraic expression for the correction done by the first perturbation
\begin{equation}
    \Ttotmax{\pert{1}(t)} -\Ttotmax{\pert{0}(t)} \approx -\frac{s n}{s n+2} \frac{\Tinf}{T_0} \chi(T_0)t .
\end{equation}
The correction is negative, meaning that the extra diffusion due to the background makes the peak decrease faster, as expected. Furthermore, it is also proportional to ${\chi(T_0) t}$ and ${\Tinf/T_0}$, which are assumed to be small in this regime. \change{In the following, $\chi$ will always mean $\chi(T_0)$, as in Eq.~\eqref{eq_chi}.}

Consider next ${\chi t \gg 1}$.
In this regime, we focus first on the solution's width.
As seen from Eq.~\eqref{eq_R_chit}, the widening to zeroth order depends on $\phi_0$.
However, this expression neglects $\phiinf$ that is illustrated in Fig.~\ref{fig_Tinf}, which corresponds to the background and increases with time as
\begin{equation}
    \phitot(t) = \phi_0 + \phiinf(t) 
    \approx  \phi_0 +  \Omega_{s} \Tinf R^{s}(t).
\end{equation}
To first order, we use the unperturbed radius $R\pert{0}$ to approximate the extra quantity $\phiinf$, which gives
\begin{equation}
    R\pert{1}(t) 
    = R\pert{0}(t) \left(\frac{\phitot}{\phi_0}\right)^{\tfrac{n}{s \, n+2}}
    \xrightarrow{\phiinf\ll\phi_0} R\pert{0}(t)\left(1 +  \frac{n G_{s,n} \Omega_s}{s \, n+2}\frac{\Tinf R^{(0)s}}{T_0 R_0^s}\right).
    \label{eq_R_pert1}
\end{equation}
The correction is positive, meaning that the background causes the distribution to widen faster.
The relative correction is proportional to ${\Tinf/T_0}$, which is small, and to ${(R\pert{0}(t)/R_0)^s}$, which gradually increases.
Note, even though the limit ${\phiinf\ll\phi_0}$ is only taken for the final right-hand side (RHS) of Eq.~\eqref{eq_R_pert1}, this first-order correction will be erroneous when ${\phiinf\gtrsim\phi_0}$. 

To make the expression in Eq.~\eqref{eq_R_pert1} approximately valid for ${\chi t\lesssim1}$, we do an approximate asymptotic matching by reintroducing the 1 in the parenthesis of Eq.~\eqref{eq_R}
\begin{equation}
    R\pert{1}(t) = R_0\left[1+ \left(\dfrac{\phitot}{\phi_0}\right)^{n} \change{\chi t} \right]^{\tfrac{1}{s \, n+2}}.
    \label{eq_R_Xt_mod}
\end{equation}

We now have the expression $\Ttotmax{\pert{1}(t)}$ for ${\chi t\ll 1}$ and $R\pert{1}(t)$ for ${\chi t \gg 1}$. To complete the pairs, we use that the total quantity above $\Tinf$ is still equal to $\phi_0$ as in Eq.~\eqref{eq_phiG}, enforced by setting ${(\Ttotmax{\pert{1}(t)}-\Tinf)(R^{(1)}(t))^s = T_0 R_0^s}$. That gives
\begin{align}
    \Ttotmax{\pert{1}(t)}
    &= \begin{cases}
        \Tinf + T_0\left[1+\left(1+\dfrac{\Tinf}{T_0}\right)^{n} \change{\chi t}
        \right]^{-\tfrac{s}{s \, n+2}}
        &\text{, if $\chi t \ll 1$}\\
        \Tinf + T_0 \left[1+\left(\dfrac{\phitot}{\phi_0}\right)^{n} \change{\chi t} \right]^{-\tfrac{s}{s \, n+2}} 
        &\text{, if $\chi t \gtrsim 1$}\\
        \rightarrow \Tinf + 
        \left( \dfrac{n}{s \, n+2}\dfrac{(G_{s,n}\phitot)^{\tfrac{2}{s}}}{2Kt} \right)^{\tfrac{s}{s \, n+2}}
        &\text{, if $\chi t \gg 1$}
    \end{cases}
    \label{eq_T_pert}
    \\
    R\pert{1}(t) 
    &= \begin{cases}
        R_0 \left[
            1+ \left(1+\dfrac{\Tinf}{T_0}\right)^n \change{\chi t}
        \right]^{\tfrac{1}{s \, n+2}}
        &\text{, if $\chi t \ll 1$}\\
        R_0\left[1+\left(\dfrac{\phitot}{\phi_0}\right)^{n} \change{\chi t}
        \right]^{\tfrac{1}{s \, n+2}} 
        &\text{, if $\chi t \gtrsim 1$} \\ 
        \rightarrow 
        \left( \tfrac{s \, n+2}{n}2K(G_{s,n}\phitot)^n t\right)^{\tfrac{1}{s \, n+2}}
        &\text{, if $\chi t \gg 1$}.
    \end{cases} 
    \label{eq_R_pert}
\end{align}
Note that only the second term of ${\Ttotmax{\pert{1}(t)}}$, not $\Tinf$, should be multiplied by ${[1-(r/R\pert{1})^2]^{1/n}}$ as in Eq.~\eqref{eq_T}.


\subsection{Validity and limitations of the first-order solution}
\label{sec_theory_validity}

The first-order solutions in Eqs.~(\ref{eq_T_pert}-\ref{eq_R_pert}) are adjustments of the zeroth-order solutions in Eqs.~\eqref{eq_R}~and~\eqref{eq_T_pert0}. 
At large times ${\chi t \gg 1}$, the adjustments stem from expanding the quantity ${\phi_0}$ by $\phiinf$, equivalent to a factor ${(1+\phiinf/\phi_0)}$.
One can imagine that a second-order solution would include an additional term of order ${(\phiinf/\phi_0)^2}$. 
For an exponent ${\alpha \lesssim 1}$, being ${\tfrac{n}{s \, n+2}<1}$ and ${\tfrac{2}{s \, n+2}<1}$ for the width and peak, respectively, the adjustment becomes
\begin{equation}
    \left(1+\frac{\phiinf}{\phi_0} + \frac{\phiinf^2}{\phi_0^2} \right)^\alpha
    = 1 + \alpha \frac{\phiinf}{\phi_0} 
        + \frac{\alpha(\alpha-1)}{2} \frac{\phiinf^2}{\phi_0^2}
        + \alpha \frac{\phiinf^2}{\phi_0^2}
        + \order{\frac{\phiinf^3}{\phi_0^3}}.
    \label{eq_errorestimate}
\end{equation}
The 2nd and 3rd terms on the RHS stem from the first-order perturbation, estimating the modification by the first-order perturbation.
The 4th term estimates the modification by the second-order perturbation. It is this term that must be small for the first-order perturbation to be a good approximation.
The 5th term of higher order reminds us that these perturbation expressions are not generally valid estimates when ${\phiinf\sim\phi_0}$.

This first-order solution also has a significant limitation. It modifies the peak and width but assumes the zeroth-order shape. 
In section~\ref{sec_theory_Tinf}, we argued that the background will modify the shape to have wider tails. Therefore, we expect an unrepresentative larger error for the boundary $R$ beyond which ${\Ttot=\Tinf}$. Instead, we will study the evolution of $R_p$, defined as the radius where the source has height equal to $p$ times the peak value above the background
\begin{equation}
    \Ttot(R_p,t) = \Tinf + p\left(\Ttotmax{(t)}-\Tinf\right).
    \label{eq_Rp}
\end{equation}
A common example of this is the half-width-half-maximum (HWHM), given by ${p=0.5}$.
Since these are self-similar shapes, both $R(t)/R(0)$ and $R_p(t)/R_p(0)$ have the same time dependence, given by Eq.~\eqref{eq_R} in the limit ${\Tinf\rightarrow 0^{+}}$.

\section{Numerical validation of perturbation theory}
\label{sec_testtheory}

{
	
    \begin{figure*}
        \renewcommand{\mysz}{0.33\textwidth}
        \centering
        \subfloat[]{\includegraphics[width=\mysz]{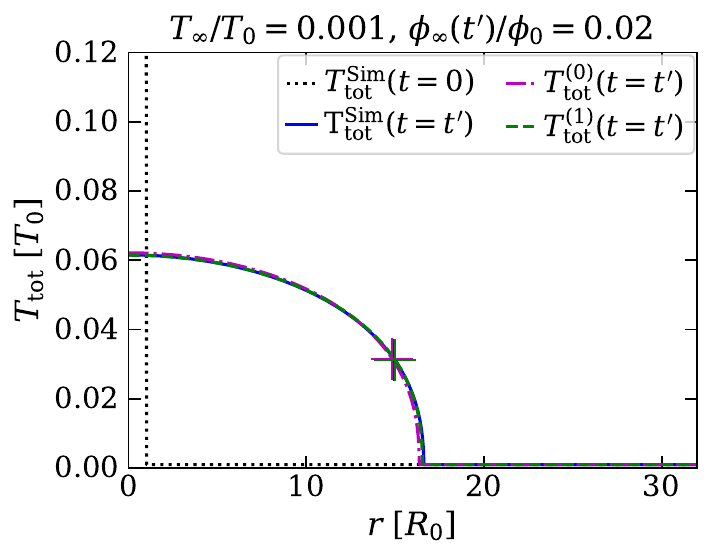}
        \label{fig_test_scana}}        
        \subfloat[]{\includegraphics[width=\mysz]{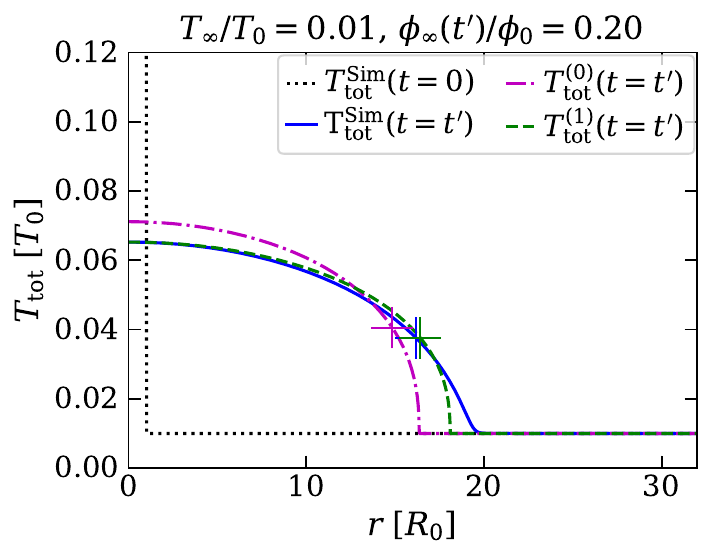} \label{fig_test_scanb}} 
        \subfloat[]{\includegraphics[width=\mysz]{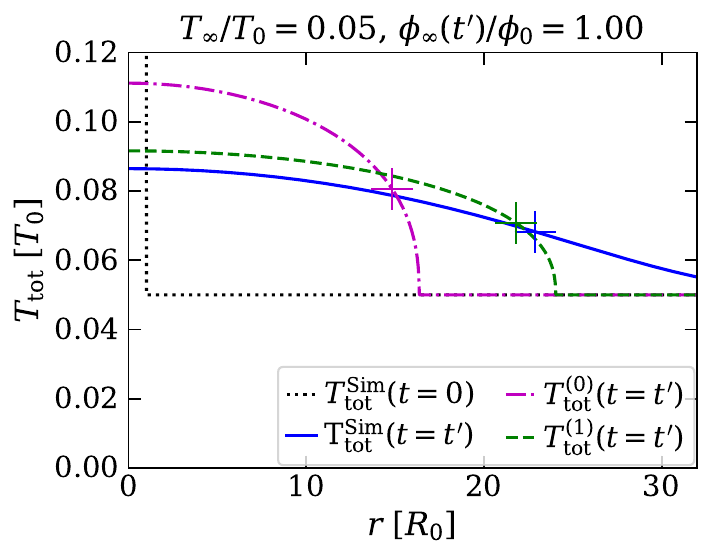} \label{fig_test_scanc}}
        \caption{Verification of analytical derivation. 
            Each plot shows the \change{IC}, as well as the simulation run using \texttt{solve\_ivp} in \texttt{Python}, zeroth-order solution $T\pert{0}$, and first-order solution $T\pert{1}$, all at a later time such that  ${\chi t \gg 1}$.
            The colored `+' corresponds to the \change{HWHM} point of each calculation. This is done for increasing background values ${\Tinf/T_0}$ given by the plot titles. The simulation box extends to ${r=100~R_0}$ in (c), preventing any impacts caused by the boundary condition. }
        \label{fig_test_scan}
    \end{figure*}

    \begin{figure}[t]
        \centering
            \includegraphics[width=0.96\columnwidth]{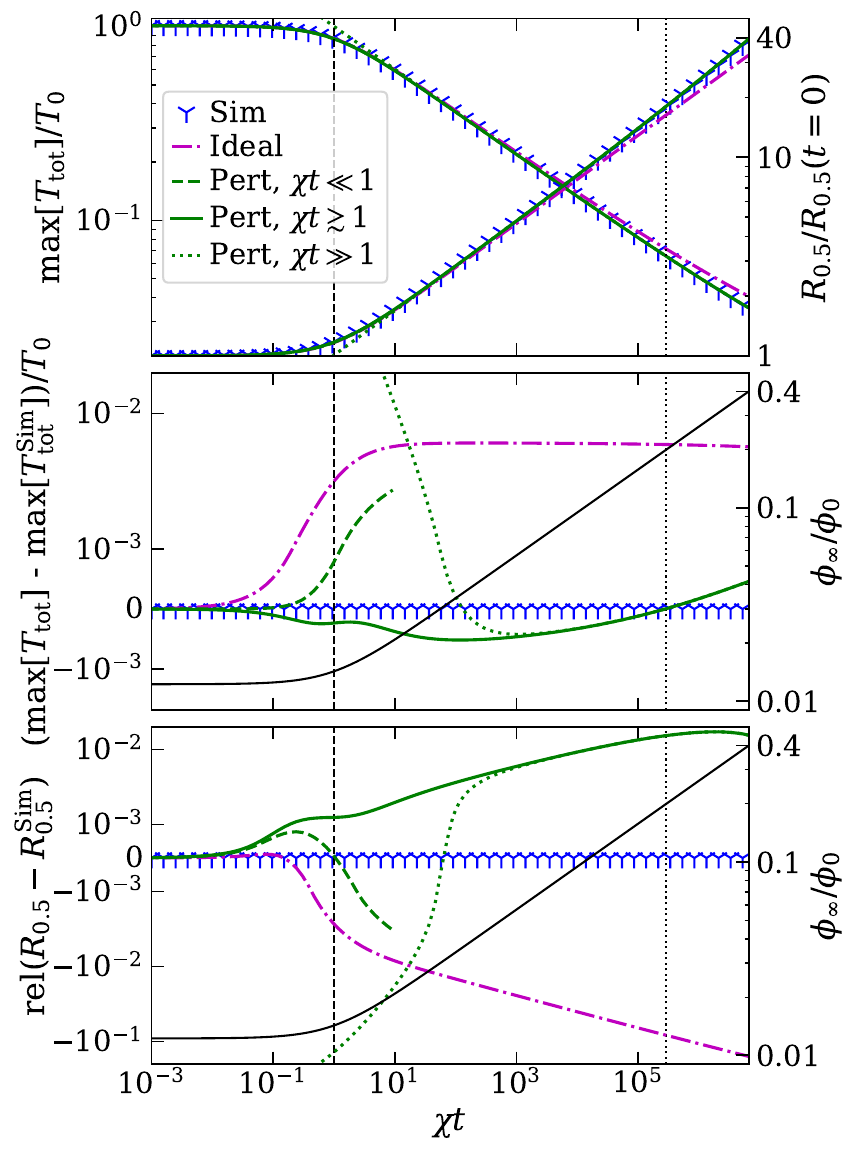}%
        \caption{
            Verification of analytical derivation for ${\Tinf/T_0=0.01}$. Comparison as a function of ${\chi t}$ for a simulation in \texttt{Python} (blue markers), as well as the zeroth-order (magenta curve) and first-order solutions (three green curves). 
            In the top panel, the absolute reduction of $\Ttotmax$ and widening of HWHM ($R_{0.5}$) is shown. 
            In the following two panels, the difference to the simulation is shown for the temperature and width, respectively. These panels also include the evolution of ${\phiinf/\phi_0}$ in black.
            The dashed vertical line marks ${t=1/\chi}$ and the dotted vertical line marks ${t=t'\gg1/\chi}$, the time of the snapshot in Fig.~\ref{fig_test_scanb}.
        }
        \label{fig_test_errorplot}
    \end{figure}
}


We have run a controlled numerical experiment to test the analytical expressions for the ideal solution $T\pert{0}$ and first-order solution $T\pert{1}$.
This is done for the normalized parameters presented in Table~\ref{table_test}, including three different values for $\Tinf$, all smaller than $T_0$. The exponent ${n=5/2}$ and dimensionality ${s=1}$ corresponds to thermal conductivity in the Solar corona, as described in section~\ref{sec_intro_TD}. 
The initial quantity is calculated from Eq.~\eqref{eq_phiG} to be ${\phi_0\approx1.64}$. 
The exponent $\alpha$ in Eq.~\eqref{eq_errorestimate} takes the values ${\tfrac{n}{s \, n+2}=\tfrac{5}{9}}$ and ${\tfrac{2}{s \, n+2}=\tfrac{4}{9}}$ for the width and peak, respectively.
\begin{table}[h!]
\caption{Parameters for numerical verification of theory}              
\label{table_test}      
\centering                                      
\begin{tabular}{l c c c c c c}          
\hline\hline                        
Parameter  & $n$   & $s$   & $K$   & $R_0$ & $T_0$ & $\Tinf/T_0$    \\
\hline                                   
Value       & $5/2$ & 1     &  1    & $1$   & $1$   & $\{0.001,0.01,0.05\}$\\    
\hline
\end{tabular}
\end{table}

The numerical results have been calculated with an implicit scheme (BDF), using \texttt{scipy.solve\_ivp} in \texttt{Python} with the \change{second}-order stencil 
\begin{equation}
    \pdv{T_{i}}{t}
    = \frac{K}{\Delta r}\left[
    \frac{T^n_{i+1}+T^n_{i}}{2}
    \frac{T_{i+1}-T_i}{\Delta r} - 
    \frac{T^n_{i}+T^n_{i-1}}{2}
    \frac{T_{i}-T_{i-1}}{\Delta r}
    \right],
\end{equation}
using ${\Delta r=R_0/250=\num{4e-3}}$. Hence, the initial peak is sufficiently well resolved to make numerical artifacts negligible.

The results of the scan are presented in Fig.~\ref{fig_test_scan}. The calculations have been run equally long, until a time $t'$ such that ${\phiinf(t')=\phi_0}$ in Fig.~\ref{fig_test_scanc}. This occurs when the zeroth-order radius is ${R\pert{0}(t')=16.35~R_0}$, which is in the regime ${\chi t \gg 1}$. Hence, it is the perturbation in the second line of Eq.~\eqref{eq_R_pert} that has been used to compare to the simulation.
For the first case, ${\Tinf/T_0=0.001}$, the final background quantity is negligible compared to the source quantity, ${\phiinf(t')/\phi_0=0.02}$. Hence, the zeroth- and first-order solutions are visually almost indistinguishable as the peak and width have been adjusted by approximately ${1\%}$. Compared to the simulation, the relative error of the peak has decreased from ${1\%}$ to ${0.1\%}$.
For the second case, ${\Tinf/T_0=0.01}$, the ideal and first-order solutions have visually diverged. It is the latter that best approximates the simulation. 
The extra quantity is still small but \change{non}negligible, ${\phiinf(t')/\phi_0=0.20}$. 
In this case, the width has increased by ${11\%}$ due to the first-order perturbation, while the estimated second-order correction in Eq.~\eqref{eq_errorestimate} is only ${2.2\%}$.
Note that the peak value $\Ttotmax{\pert{1}(t)}$ and HWHM are almost identical to the simulation. However, the expected slightly heavier tail in the simulation is visible. 
For the third case, ${\Tinf/T_0=0.05}$, the extra quantity has reached equality to the source quantity, ${\phiinf(t')=\phi_0}$. Even though the first-order solution is better than the ideal case, it is no longer a good approximation of the simulation. Both the peak and HWHM are visually different, and the shape is even more different. This was expected, as second- and higher-order terms of ${(\phiinf(t')/\phi_0)=1}$, will no longer be negligible. 

We have studied the time evolution in greater detail for the intermediate background value, ${\Tinf/T_0=0.01}$. The evolution of $\Tmax$ and the HWHM ($R_{0.5}$) are presented in Fig.~\ref{fig_test_errorplot}. 
The absolute value curves in the first panel are visually indistinguishable, except the ideal zeroth-order solution that deviates \change{toward} the end with a too narrow and peaked distribution, as expected. 
After ${\chi t =10}$, the peak error levels out at almost ${6\times10^{-3}\,T_0}$. It does not grow more for these parameters because the simulation is slower for a smaller peak and both peaks decrease \change{toward} ${\Ttotmax{}\rightarrow \Tinf=10^{-2}\,T_0}$. 
However, the relative error in width ends at ${-16\%}$.

The first-order perturbation has three different curves, depending on the value of ${\chi t}$. They are given in the same order by the legend as by the three lines in Eqs.~(\ref{eq_T_pert}-\ref{eq_R_pert}). The third line only coincides with the second for ${\chi t > 10^{3}}$. This delayed convergence supports the need for the approximate matching introduced in Eq.~\eqref{eq_R_Xt_mod}. The second line also coincides rather well with the first line for ${\chi t < 0.1}$, and thereby connecting the two regimes. Therefore, the second expression will be used in the following sections. 

The first-order perturbation reduces the differences between the simulation and the ideal zeroth-order solution up to the end of the simulation when ${\phiinf/\phi_0=0.4}$. 
The absolute relative peak error maxes at ${\num{5e-4}\, T_0}$, an order of magnitude smaller than before.
The relative width error has also dropped by an order of magnitude, to ${1.6\%}$ at the end of the simulation, notably now with the opposite sign.


The ideal zeroth-order solution appears to better estimate the width at small times ${\chi t< 0.3}$, as seen in the bottom panel of Fig.~\ref{fig_test_errorplot}.
First of all, note that this small discrepancy is relative to the ideal HWHM, which is initially small as well. 
The relative mismatch decreased by reducing the grid spacing ${\Delta r}$ to the current choice, as the singular point at ${r=R}$ is difficult to resolve with a uniform grid.
Here, the absolute difference to the first-order perturbation is approximately ${\Delta r/4}$.
Secondly, note that the distribution shape is expected to adjust during this time regime, ${\chi t \ll 1}$. 
By close inspection, the discrepancy to the first-order perturbation is caused by the marginally heavier tails in the simulation, making the width at half maximum marginally narrower. This effect is illustrated by comparing the curves for ${n=\tfrac{5}{2}}$ and, \change{for example}, ${n=0}$ in Fig.~\ref{fig_selfsim_n}.

\section{Benchmarking a numerical code}
\label{sec_benchmark}

\subsection{Designing a numerical test}
\label{sec_test_design}

The previous sections illustrate several requirements to remember when designing a test for your \change{non}linear diffusion solver for any exponent $n$. 
\begin{enumerate}
    \item   $\alpha(\phiinf/\phi_0)^1\ll1$, if zeroth-order solution\\
            $\alpha(\phiinf/\phi_0)^2\ll1$, if first-order solution\\
            where ${\alpha=\max\left(\tfrac{n}{s \, n+2},\tfrac{2}{s \, n+2}\right)}$
    \item ${\chi t\gg 1}$
    \item $R(t) \gtrsim R\pert{0}(t) = R_0(1+\chi t)^{\tfrac{1}{s \, n+2}} \gg R_0 $ 
    \item ${R_0\gg \Delta r}$
\end{enumerate}
Point~1 states that one must use the zeroth and first-order solutions only in their regions of validity. This sets an upper limit for the time. 
Point~2 incorporates that one additionally must wait until the initial distribution has adjusted to the approximately self-similar solution, setting as well a lower limit for time. 
Obviously, the lower limit for time must be lower than the upper limit.
When it comes to spatial discretization, the lower limit on time from point~2 also sets a lower limit for how much the distribution has to widen, given by point~3.
Point~4 includes additionally that the initial distribution must be sufficiently resolved numerically. 
The latter two points combined make a lower requirement for how many grid points one needs.
A functioning set of parameters for this is given by the parameters in Table~\ref{table_test}, 
with ${\Tinf/T_0\leq0.01}$ 
and ${R_0/\Delta r\geq 10}$, 
run until a time ${t'}$ such that ${R\pert{0}(t')/R_0\geq2}$, but also ${\phiinf(t')/\phi_0\leq0.10}$. 
This can be both a one-off test, as exemplified below, and continuous integration in the form of a test to be automatically run at each new commit to a version control software such as \texttt{git}.

{ 
\begin{figure*}
    \centering
    \subfloat[]{\includegraphics[width=.33\linewidth]{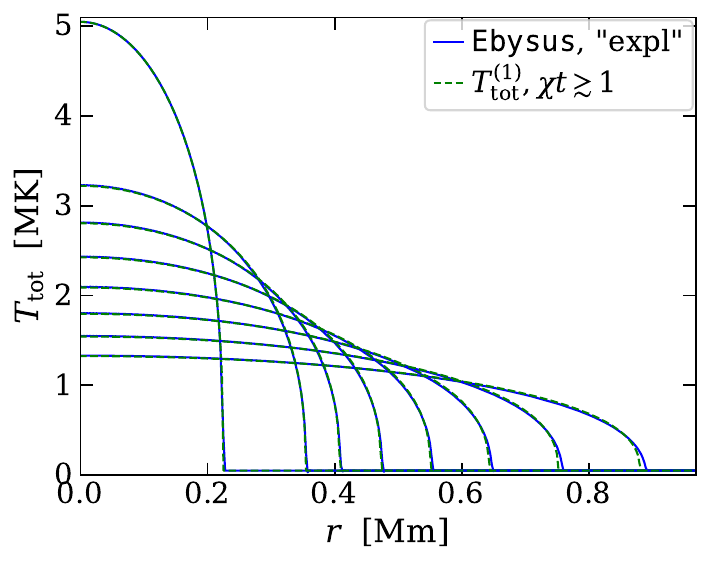}}%
    \subfloat[]{\includegraphics[width=.33\linewidth]{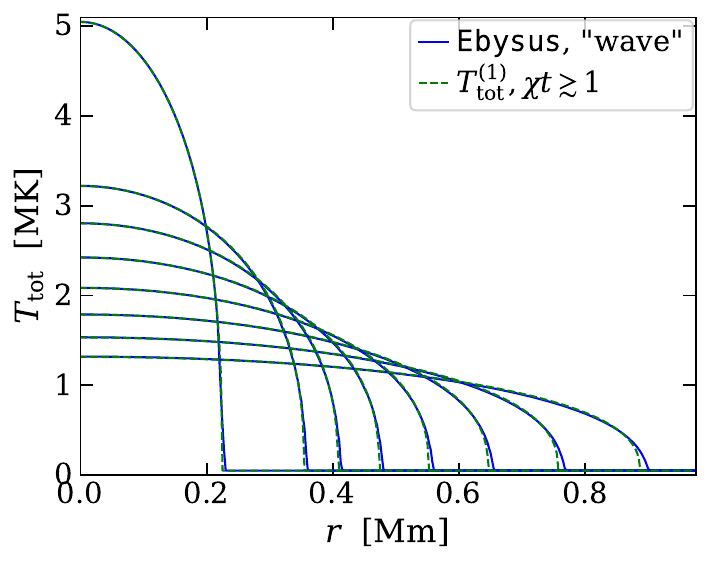}}
    \subfloat[]{\includegraphics[width=.33\linewidth]{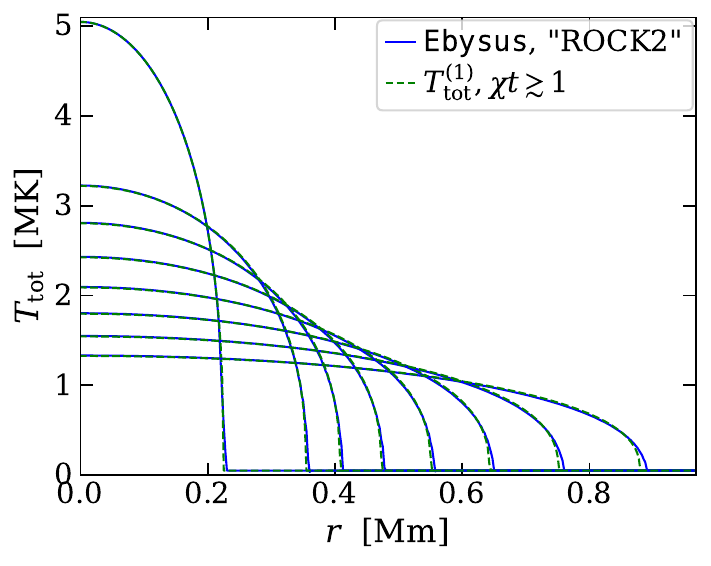}}
    \caption{Comparison between \ebysus{} and the first-order expression in Eqs.~(\ref{eq_T_pert}-\ref{eq_R_pert}). The narrowest curve is the \change{IC}, and the following curves are given after exponentially longer time as {[3~s, 6~s, 12~s, $\dots$]}.
    \ebysus{} has been run with different Spitzer methods in the different panels:
        (a) is explicit, 
        (b) is the wave method from~\citet{rempel2016},
        and (c) is the ROCK2 method~\citep{abdulle_rock2_2001}.
	}
	\label{fig_ebysus}
\end{figure*}

\begin{figure}[t]
    \centering
    \includegraphics[width=0.96\columnwidth]{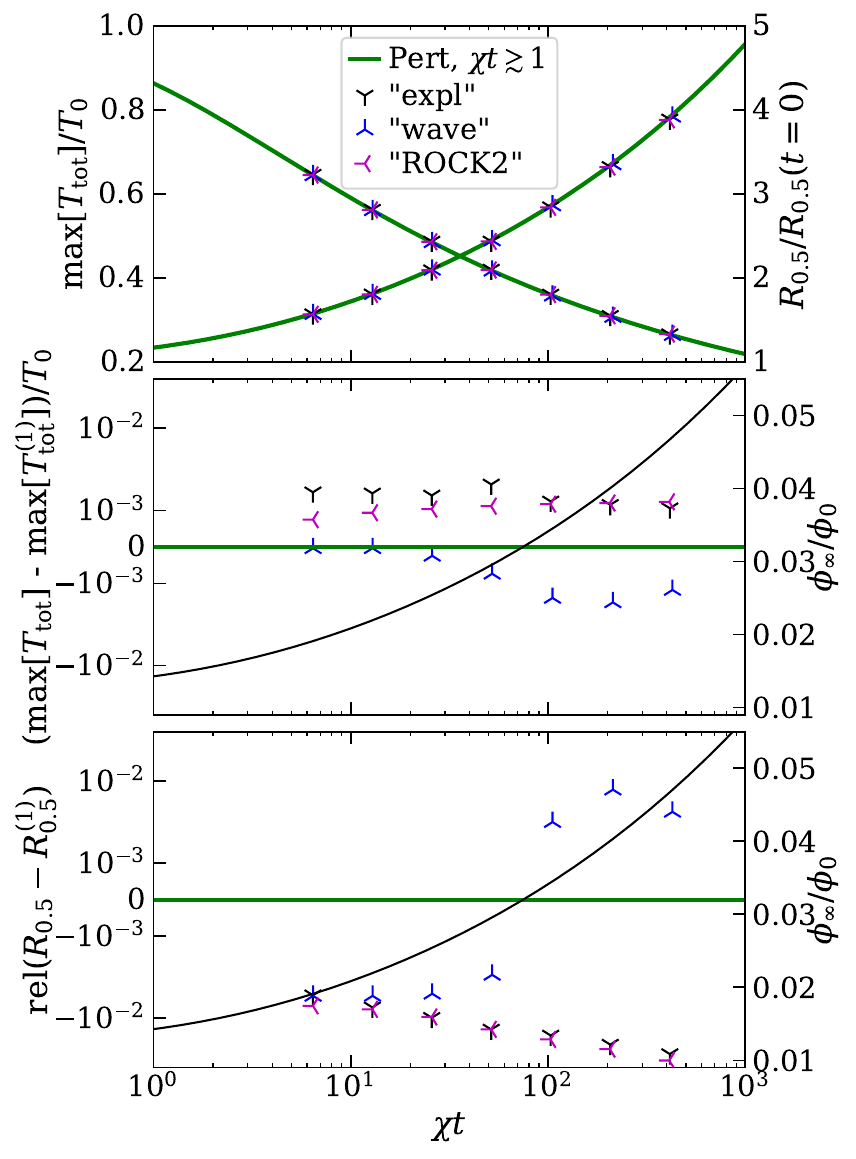}
    \caption{
        Error plot for the test of \ebysus{} presented in Fig.~\ref{fig_ebysus}. See the description in the similar Fig.~\ref{fig_test_errorplot}. Note that, here, several $y$~axes are linear and 
        the error is shown relative to the first-order perturbation theory. The symmetric logarithmic error scales in the two bottom panels go from $\num{-4e-2}$ to $\num{4e-2}$.
        }
    \label{fig_errorplot_ebysus}
\end{figure}
}

\subsection{Test of Spitzer conductivity in \ebysus{}} 
\label{sec_test_ebysus}


The theory described in this paper has been used to benchmark various solvers for Spitzer conductivity implemented in~\ebysus{}. The benchmark has been done for a one-off case with parameters comparable to those in the Solar atmosphere. 
The electron density is ${N=\SI{1e12}{\per\centi\meter\cubed}}$ and the conductivity coefficient is set to ${\kappa_\parallel^\ast=\SI{1.1e-6}{\erg\per\second\per\cm\per\kelvin\tothe{7/2}}}$, which is realistic for the Solar atmosphere~\citep{braginskii1965,spitzer1962}.
The background temperature combined with the released energy gives an initial temperature profile as in Eq.~\eqref{eq_T_pert0} with ${\Tinf=\SI{5e4}{\kelvin}}$, ${T_0=\SI{5e6}{\kelvin}}$, and ${R_0=\SI{225}{\kilo\meter}}$.

The comparison to the first-order solution is given as a function of the radius in Fig.~\ref{fig_ebysus}. The agreement is clear. \change{Toward} the end, the tails are slightly wider in the simulations compared to the theory, as expected from theoretical considerations in Sec.~\ref{sec_theory_Tinf} and also seen in Fig.~\ref{fig_test_scan}.

The corresponding errors of the peak and HWHM are given in Fig.~\ref{fig_errorplot_ebysus}. All three methods agree reasonably well, the wave method being slightly off compared to the other two. The error seems to grow with time, especially in radius. Since the first-order solution showed a lower peak value and larger HWHM than the trusted simulation in Fig.~\ref{fig_test_errorplot}, one can argue that the explicit and ROCK2 methods are more accurate than the wave method. 
This is understandable since the wave method approximates the problem by solving a hyperbolic diffusion equation~\citep{rempel2016}.

To advance the simulations $\SI{3}{\second}$ from the IC to the second curve in Fig.~\ref{fig_ebysus} or first point in Fig.~\ref{fig_errorplot_ebysus}, 
the standard explicit method used 28656~time steps, 
the wave method used 7111~time steps, 
while the ROCK2 method uses 2699~time steps. 
The length of the time steps was calculated dynamically to ensure numerical stability, depending on the diffusivity $D(T)$ and the spatial resolution. These numbers exemplify that many time steps are required to solve \change{non}linear diffusion and that great \change{speedup} can be achieved by seeking alternatives to the standard explicit method, such as the competitive wave and ROCK2 methods.
Q.~Wargnier et al. 2024 (in prep.) will describe further details on the \ebysus{} benchmark using both ROCK2 and PIROCK~\citep{abdulle2013}.
Furthermore, G.~Cherry et al. (in prep.) will use the theory derived here to study the Spitzer conductivity modeling in the \bifrost{} code, aiming to analyze the efficacy of the different numerical methods.

It is important to verify that these tests followed the benchmark requirements given in Sec.~\ref{sec_test_design}.
The final distribution is well within the first requirement, with ${\phiinf/\phi_0=0.05\ll1}$ and ${\alpha\in\{5/9,4/9\}}$.
The curves shown, except for the IC, are for ${t\geq\SI{3.0}{\second}\gg1/\chi=\SI{0.47}{\second}}$. As a consequence, their widths ${R(t>0)}$ are significantly larger than $R_0$. Even larger widths could have been necessary if not for the choice of setting the IC equal to the relevant self-similar solution. There is no concrete estimate for when a general source distribution reaches the self-similar solutions, even in the ideal case.
Lastly, the IC is well resolved, with resolution such that ${R_0/\Delta r=100\gg1}$.

\section{Nanoflare experiment}
\label{sec_flares}

Finally, we will analyze the thermal Spitzer conductivity during nanoflares in the Solar atmosphere in isolation from other physical processes. We are inspired by the studies of~\citet{polito2018} and~\citet{testa2014}. Here, we will focus on the impact of only conductivity without radiation, \change{nonequilibrium} ionization effects, advection, or electron beams.
A total of 10 configurations have been analyzed, consisting of 1 reference model and 9 variations where we study the impact of changing different key parameters focusing \change{solely} on thermal conduction.

\subsection{Model setup}
\label{sec_flare_model}

A reference experiment has been constructed similar to the thermal conduction experiment in~\citet{polito2018} with an initial peak temperature of~$\SI{1}{\mega\kelvin}$ and coronal electron density of ${N\lesssim\SI{1e9}{\per\cm\cubed}}$.
They increased the energy by ${\dot{Q}=\SI{6e23}{\erg\per\second}}$ for ${\SI{10}{\second}}$ over an area of ${A=\SI{5e14}{\centi\meter^{2}}}$ (corresponding to a diameter of $\SI{0.25}{\mega\meter}$) transverse to the magnetic field and over a length of $\SI{9}{\mega\meter}$ along a magnetic loop. 
This is appropriate for a nanoflare according to the work of~\citet{testa2014} and the description of nanoflares by~\citet{parker1988}. 
The released energy increased the coronal temperature to ${\SI{20}{\mega\kelvin}}$ in a few seconds. The energy was conducted down to the transition region, which caused the denser plasma to move quickly up into the corona.

Our study focuses on the importance of thermal conduction at the beginning of a nanoflare, and how it depends on the key parameters. Spitzer's conductivity coefficient is again set to ${\kappaeff=\SI{1.1e-6}{\erg\per\second\per\cm\per\kelvin\tothe{7/2}}}$~\citep{braginskii1965,spitzer1962}.
The IC of the reference experiment is consistent with an immediate energy increase equivalent to a typical nanoflare energy release over $\SI{1}{\second}$, ${E^\mathrm{ref}=\dot{Q}\times\SI{1}{\second}=\SI{6e23}{\erg}}$. All the energy is unrealistically assumed to increase the temperature of the plasma, no energy goes into macroscopic kinetic energy or ionization.
The energy is released over an area ${A=\SI{5e14}{\centi\meter^{2}}}$ transverse to the magnetic field and over a self-similar shape with radius ${R_0^\mathrm{ref}=\SI{0.15}{\mega\meter}}$ parallel to the magnetic field, corresponding to an effective diameter of $\SI{0.25}{\mega\meter}$. 
The reference electron density is set to ${N^\mathrm{ref}=\SI{1e9}{\per\centi\meter\cubed}}$.
Assuming an ideal gas, that gives an initial peak temperature of ${T^\mathrm{ref}_0=\SI{140}{\mega\kelvin}}$ that quickly diffuses over the background temperature of ${\Tinf^\mathrm{ref}=\SI{1}{\mega\kelvin}}$. Even though this peak temperature is unrealistic, we choose such a large $T_0^\mathrm{ref}$ and correspondingly small $R_0^\mathrm{ref}$ to illustrate the impact of different key parameters better.

In addition to the reference experiment, several small parameter scans have been performed around the reference parameters $(T_0^\mathrm{ref},\Tinf^\mathrm{ref},E^\mathrm{ref},R_0^\mathrm{ref},N^\mathrm{ref})$
\begin{enumerate}
    \item Change $T_0$ and $R_0\propto1/T_0$ with $E$ unchanged.
    \item Change $T_0$ and $E\propto T_0$ with $R_0$ unchanged.
    \item Change only $\Tinf$.
    \item Change $N$ with $E$ unchanged, which in turn changes both ${T_0\propto1/N}$ and the ${K=\kappaeff/c_v\rho\propto1/N}$ in Eq.~\eqref{eq_dTdt_cond_intro}.
\end{enumerate}

\subsection{Results}
\label{sec_flare_results}

\begin{figure}[t]
    \centering
    \renewcommand{\mysz}{0.93}
    \subfloat{\includegraphics[width=\mysz\linewidth,trim=0 2mm 0 0mm, clip]{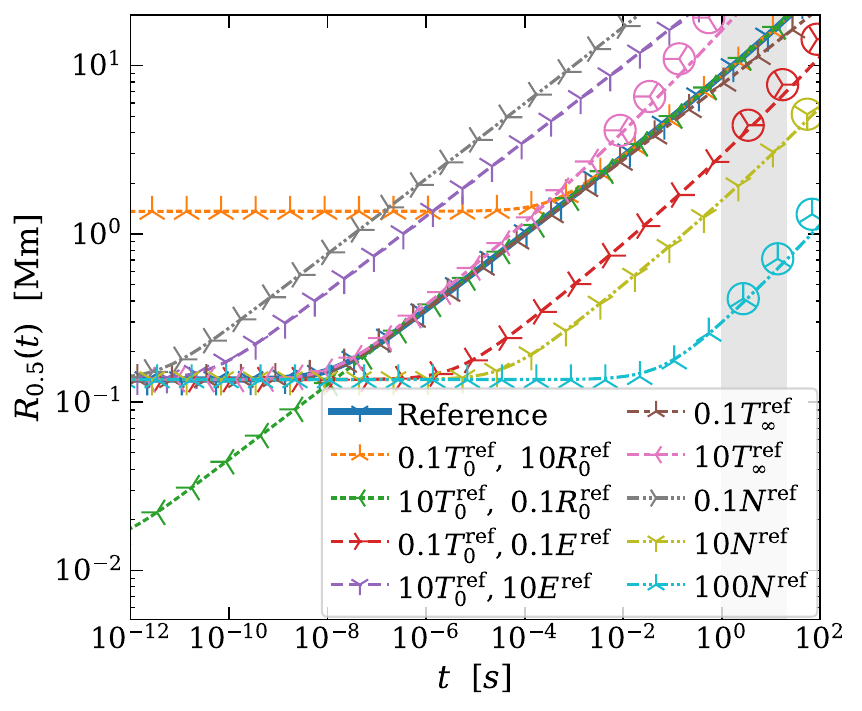}}\\
    \vspace{-1em}
    \subfloat{\includegraphics[width=\mysz\linewidth,trim=0 2mm 0 0mm, clip]{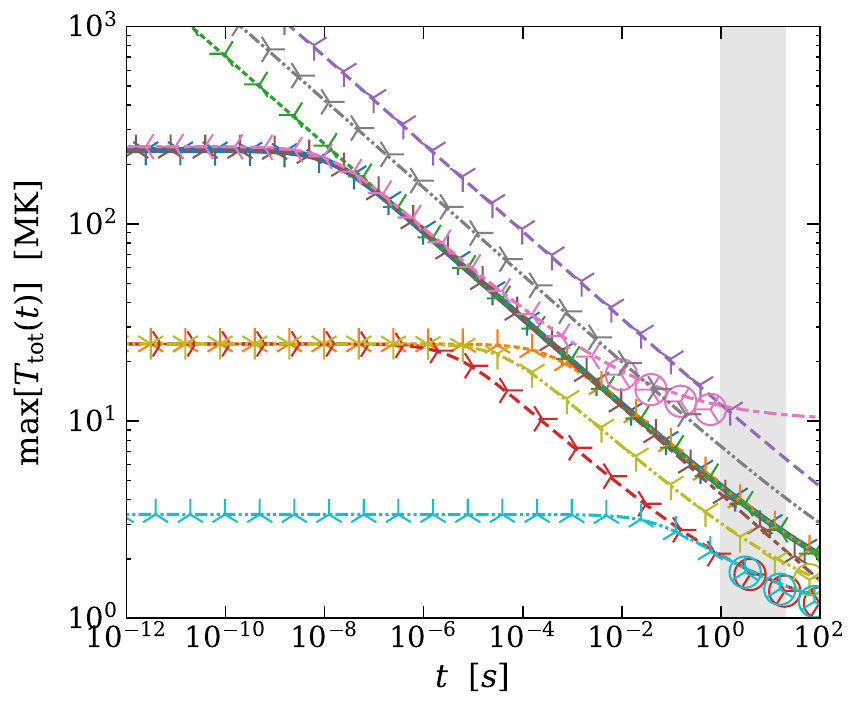}}
    \caption{
        Evolution of the width and maximum of the temperature peak after the concentrated release of energy corresponding to $\SI{1}{\second}$ of a nanoflare in the Solar corona. 
        The model parameters are detailed in Sec.~\ref{sec_flare_model}.
        The top panel shows the HWHM, while the bottom panel shows maximum temperature.
        The curves are calculated with Eqs.~(\ref{eq_T_pert}-\ref{eq_R_pert}) and the points are calculated with simulations. 
        Where the points are encircled ${\phiinf\geq\phi_0}$, making the equations a poor approximation.
        The \change{gray} shaded area corresponds to ${t\in[\SI{1}{\second},\SI{20}{\second}]}$, the time frame of the simulations in~\citet{polito2018}.
    }
    \label{fig_paramscan}
\end{figure}

The numerical experiments have been simulated with the code described in Sec.~\ref{sec_testtheory} with ${R_0/\Delta r\geq 25}$ and evaluated with the first-order theory described in Eqs.~(\ref{eq_T_pert}-\ref{eq_R_pert}). 
The evolution of the HWHM $R_{0.5}$ (radius at $50\%$ of maximum) and maximum temperature $\Ttotmax{}$ are presented in Fig.~\ref{fig_paramscan}. 
The theoretical curves agree well with the simulations in most cases. For the encircled points, marking when ${\phiinf\geq\phi_0}$, the agreement becomes worse, as expected. This is due to the heavier tails illustrated in Fig.~\ref{fig_test_scanc} that are not properly represented outside the region of validity of the first-order solution. The general behavior is, nevertheless, well included in the analytical first-order solutions.

The reference simulation shows the behavior due to the conduction of energy released in the first second of a nanoflare in the corona. Within ${\SI{1}{\second}}$, the energy has been spread out over an area with HWHM of approximately $\SI{9}{\mega\meter}$, or full width of $\SI{18}{\mega\meter}$, and the maximum temperature has dropped to ${\SI{2}{\mega\kelvin}=2\,\Tinf}$. Hence, the extent of the coronal heating used in~\citet{polito2018} could have been achieved by the conduction alone within a second.

When changing the parameters from the reference configuration, several important results are prominent, many of which can be understood from the analytical solution.
Changing only $T_0$ and $R_0$ affects only the early phase of the evolution. At ${\chi t\gg1}$, these curves converge to the reference configuration. Hence, the spatial extent of the IC parallel to the magnetic field is of little physical importance for long-term conductivity.

When increasing $T_0$ and $E$, while keeping $R_0$ fixed, it causes a larger $\chi$ in Eq.~\eqref{eq_chi} and thereby an earlier evolution since ${\chi t =1}$ occurs for a correspondingly earlier $t$. Hence, the energy of a more energetic event will be conducted faster.
Changing the released energy by an order of magnitude up (down), as in the scan, changes both the radius and the temperature above the background (${\Ttotmax{}-\Tinf}$) after $\SI{1}{\second}$ by a factor of ${\sim\!\!\sqrt{10}\approx3}$ up (down).
Note that this change of $E$ along the field line can also be due to a change of area $A$ perpendicular to the field.

Increasing only the background temperature $\Tinf$ makes the widening occur slightly faster \change{toward} the end, as seen by the pink curves and points relative to the reference. That is because, \change{toward} the end of the simulation, the max temperature approaches ${\Tinf=10^7~\si{\kelvin}}$, and the background temperature will contribute to significant diffusion. Put differently, the first-order perturbation is a poor approximation for this case, as seen in the encircled pink points relative to the pink curves.
Here, the first-order theory will underestimate the simulated widening as ${\phiinf\geq\phi_0}$
This is the same effect as in Fig.~\ref{fig_test_scanc}.

Increasing the electron density $N$ increases the heat capacity of the plasma, thereby reducing the initial increase of temperature $T_0$ and the conductivity $K$. 
Hence, the first-order theory will be less accurate, since ${\phiinf/\phi_0}$ starts out lower.
In addition, the widening will be slower. Increasing the electron density by an order of magnitude reduces the width after $\SI{1}{\second}$ by a factor of ${\sim5.5}$.
This is one reason why the conduction slows down when it reaches the denser transition region in~\citet{polito2018}.

\section{Conclusion}


In this paper, we have stressed that every single physics module in numerical \change{multi}physics simulations must be tested thoroughly and separately. Many tests exist, but we found no appropriate test for Spitzer thermal conductivity in the Solar atmosphere. 
Therefore, we have derived an analytical first-order solution for \change{non}linear diffusivity ${D=KT^n}$ with any exponent ${n>0}$ in ${s=\{1,2,3\}}$ dimensions. Since the derivation and argumentation are general, the solution can easily be applied to other \change{non}linear diffusion problems.

The analytical solution is based on the self-similar shapes by~\citet{Pattle1959}. However, where those shapes required the source quantity to diffuse in a vacuum, the new first-order solutions allow for a finite background quantity $\Tinf$. That was a key requirement for our use case, as the temperature in the Solar atmosphere is \change{nonzero}.
In the limit ${\Tinf\rightarrow0^{+}}$, the first-order solutions coincide with the original zeroth-order solutions. 
The region of validity of both the zeroth-order and first-order solutions have been derived analytically and tested numerically.

We have proposed 4 requirements for making a benchmark based on the first-order solution: 
(i) The diffusing quantity $\phi_0$ must be large compared to the background quantity underneath it;
(ii) the simulation must be run for a sufficiently long time;
(iii) the width of the diffusing quantity must become large compared to the initial extent;
(iv) the initial source quantity must be well-resolved numerically.

Following the requirements for making a test, we have benchmarked various solvers for Spitzer conductivity in the single- and \change{multi}fluid radiative MHD code~\ebysus{}. 
They agree well with the first-order solutions. 
Going forward, 
Q.~Wargnier et al. 2024 (in prep.) will describe further the use of ROCK2 and PIROCK in the \ebysus{} code, while G.~Cherry et al. (in prep.) will test further the solvers for Spitzer conductivity implemented in the \bifrost{} code.

Finally, based on the theoretically and numerically developed understanding of Spitzer conductivity, we have analyzed its role during the start of a nanoflare event in the Solar atmosphere. 
We found that conductivity alone can spread the released energy of a representative nanoflare $\SI{9}{\mega\meter}$ in $\SI{1}{\second}$ in a representative coronal atmosphere. 
Combining our first-order derivation with a parameter scan allowed us to understand better the thermal conduction evolution in terms of the background temperature, coronal electron density, nanoflare radius, and nanoflare energy release per area perpendicular to the magnetic field. We found, as in~\citet{polito2018}, that the \change{IC} before the nanoflare release impacts the thermal conduction significantly. Particularly the electron density is crucial because it is proportional to the plasma's total heat capacity and thus inversely proportional to both the initial temperature increase and the effective temperature diffusivity $K$. The conduction slows down for either a larger electron density or smaller nanoflare energy release.

\begin{acknowledgements}

    This research has been supported by the European Research Council through the Synergy Grant number 810218 (“The Whole Sun”, ERC-2018-SyG),
    from the European Union's Horizon 2020 research and innovation programme under the Marie Skłodowska-Curie grant agreement Nº 945371,
    and by the Research Council of Norway through its Centres of Excellence scheme, project number 262622 (Rosseland Centre for Solar Physics -- RoCS).

    We would like to thank F.~Moreno-Insertis for the positive and fruitful discussions during the Whole Sun meeting in March, 2024.
      
\end{acknowledgements}



\bibliographystyle{aa} 
\bibliography{TTD_ref} 

\begin{thebibliography}{23}
\expandafter\ifx\csname natexlab\endcsname\relax\def\natexlab#1{#1}\fi

\bibitem[{Abdulle(2002)}]{Abdulle2002}
Abdulle, A. 2002, SIAM Journal on Scientific Computing, 23, 2041

\bibitem[{Abdulle \& Li(2008)}]{abdulle_2008}
Abdulle, A. \& Li, T. 2008, Communications in Mathematical Sciences, 6, 845

\bibitem[{Abdulle \& Medovikov(2001)}]{abdulle_rock2_2001}
Abdulle, A. \& Medovikov, A.~A. 2001, Numerische Mathematik, 90, 1

\bibitem[{Abdulle \& Vilmart(2013)}]{abdulle2013}
Abdulle, A. \& Vilmart, G. 2013, Journal of Computational Physics, 242, 869

\bibitem[{{Abramowitz} \& {Stegun}(1965)}]{abramowitz_1965}
{Abramowitz}, M. \& {Stegun}, I.~A. 1965, {Handbook of mathematical functions
  with formulas, graphs, and mathematical tables} ({Dover Publications, Inc.,
  New York})

\bibitem[{Bakke {et~al.}(2022)Bakke, Carlsson, Voort, Gudiksen, Polito, Testa,
  \& Pontieu}]{bakke2022}
Bakke, H., Carlsson, M., Voort, L. R. v.~d., {et~al.} 2022, Astronomy \&
  Astrophysics, 659, A186

\bibitem[{Braginskii(1965)}]{braginskii1965}
Braginskii, S. 1965, Reviews of Plasma Physics, 1, {p.~205}

\bibitem[{Diez {et~al.}(1992)Diez, Gratton, \& Minotti}]{diez1992}
Diez, J.~A., Gratton, J., \& Minotti, F. 1992, Quarterly of Applied
  Mathematics, 50, 401

\bibitem[{Gudiksen {et~al.}(2011)Gudiksen, Carlsson, Hansteen, Hayek,
  Leenaarts, \& Martínez-Sykora}]{gudiksen2011}
Gudiksen, B.~V., Carlsson, M., Hansteen, V.~H., {et~al.} 2011, Astronomy \&
  Astrophysics, 531, A154

\bibitem[{{Kowalski} {et~al.}(2024){Kowalski}, {Allred}, \&
  {Carlsson}}]{kowalski_stellarflare_2024}
{Kowalski}, A.~F., {Allred}, J.~C., \& {Carlsson}, M. 2024, \apj, 969, 121

\bibitem[{Martínez-Sykora {et~al.}(2020)Martínez-Sykora, Szydlarski,
  Hansteen, \& Pontieu}]{sykora_2020}
Martínez-Sykora, J., Szydlarski, M., Hansteen, V.~H., \& Pontieu, B.~D. 2020,
  The Astrophysical Journal, 900, 101

\bibitem[{Moreno-Insertis {et~al.}(2022)Moreno-Insertis, Nóbrega-Siverio,
  Priest, \& Hood}]{moreno2022}
Moreno-Insertis, F., Nóbrega-Siverio, D., Priest, E.~R., \& Hood, A.~W. 2022,
  Astronomy \& Astrophysics, 662, A42

\bibitem[{{Parker}(1988)}]{parker1988}
{Parker}, E.~N. 1988, \apj, 330, 474

\bibitem[{Pattle(1959)}]{Pattle1959}
Pattle, R.~E. 1959, Quarterly Journal of Mechanics and Applied Mathematics, 12,
  407

\bibitem[{Polito {et~al.}(2018)Polito, Testa, Allred, Pontieu, Carlsson,
  Pereira, Gošić, \& Reale}]{polito2018}
Polito, V., Testa, P., Allred, J., {et~al.} 2018, ApJ, 856, 178

\bibitem[{Press {et~al.}(2007)Press, Teukolsky, Vetterling, \&
  Flannery}]{numrec2007}
Press, W.~H., Teukolsky, S.~A., Vetterling, W.~T., \& Flannery, B.~P. 2007,
  Numerical Recipes 3rd Edition: The Art of Scientific Computing, 3rd edn.
  (Cambridge University Press)

\bibitem[{Priest(1984)}]{priest1984}
Priest, E.~R. 1984, Solar Magnetohydrodynamics (D. Reidel Publishing Company)

\bibitem[{{Reep} {et~al.}(2015){Reep}, {Bradshaw}, \& {Alexander}}]{Reep_2015}
{Reep}, J.~W., {Bradshaw}, S.~J., \& {Alexander}, D. 2015, \apj, 808, 177

\bibitem[{Rempel(2016)}]{rempel2016}
Rempel, M. 2016, The Astrophysical Journal, 834, 10

\bibitem[{{Sod}(1978)}]{sod1978}
{Sod}, G.~A. 1978, Journal of Computational Physics, 27, 1

\bibitem[{Spitzer(1962)}]{spitzer1962}
Spitzer, L. 1962, Physics of Fully Ionized Gases (Interscience, New York)

\bibitem[{Testa {et~al.}(2014)Testa, De~Pontieu, Allred, Carlsson, Reale, Daw,
  Hansteen, Martinez-Sykora, Liu, DeLuca, Golub, McKillop, Reeves, Saar, Tian,
  Lemen, Title, Boerner, Hurlburt, Tarbell, Wuelser, Kleint, Kankelborg, \&
  Jaeggli}]{testa2014}
Testa, P., De~Pontieu, B., Allred, J., {et~al.} 2014, Science, 346, 1255724

\bibitem[{Zbinden(2011)}]{Zbinden2011}
Zbinden, C.~J. 2011, SIAM J. Sci. Comput., 33, 1707

\end{thebibliography}

\begin{appendix} 
%

\section{Thermal conductivity in a plasma}
\label{app_thermalconductivity}

The conductive part of the energy equation can be written as 
\begin{equation}
    \left(\pdv{e}{t}\right)_\mathrm{cond} = -\grad \cdot \textbf{F}_c, 
    \label{eq_dedt_cond}
\end{equation}
where $e$ is the internal energy per unit volume and $\textbf{F}_c$ is the heat flux vector. In a magnetized plasma, the heat flux vector can be split into two parts ~\citep[see][p.~86]{priest1984}
\begin{equation}
    \textbf{F}_c 
    =	-\grad \cdot (\kappa \grad T)
    =   -\grad_\parallel \cdot (\kappa_\parallel\grad_\parallel T) 
    	 -\grad_\perp \cdot (\kappa_\perp\grad_\perp T) ,
\end{equation}
where $\mathbf{\kappa}$ is the thermal conduction tensor and the  subscripts $\parallel$ and $\perp$ signify components parallel and perpendicular to the magnetic field vector $\textbf{B}$, respectively.
In the solar atmosphere, the perpendicular conduction is typically significantly smaller than the parallel. The parallel conduction coefficient for a fully \change{ionized} hydrogen plasma is \citep{spitzer1962}
\begin{equation}
    \kappa_\parallel = 
    \num{1.8e-5}
    \frac{T^{5/2}}{\ln{\Lambda}}
    \si{\erg\per\second\per\cm\per\kelvin}=\kappa_\parallel^\ast T^{5/2},
\end{equation}
which is close to ${10^{-6} T^{5/2}}$ in the chromosphere and corona for a fully ionized hydrogen gas~\citep{braginskii1965,priest1984}.
This ${\kappa_\parallel^\ast}$ is identical to the ${\sigma=10^{-6}~\si[per-mode=symbol]{\erg\per\second\per\cm\per\kelvin}}$ in \citet{rempel2016}.

For an ideal polytropic gas, the internal energy is related to the temperature as 
\begin{equation}
    e=c_v \rho T = \frac{k_b\rho}{\mu m_p (\gamma-1)}T ,
\end{equation}
where $c_v$ is the specific heat capacity per mass and $\mu$ the mean molecular mass.
Assuming $c_v$ to be constant and using the continuity equation, Eq.~\eqref{eq_dedt_cond} can be written on the form
\begin{equation}
\begin{split}
    \pdv{T}{t} 
    &= \frac{1}{c_v \rho}\pdv{e}{t} - \frac{T}{\rho} \pdv{\rho}{t} \\
    &= -\frac{1}{c_v \rho} \grad \cdot \textbf{F}_c
       + \frac{T}{\rho} \div (\rho \textbf{v}),
\end{split}
\label{eq_dTdt}
\end{equation}
where the RHS of the temperature evolution is split into a conductive term first and a convective term second. 
If we further assume 
a negligible perpendicular heat conduction ${\kappa_\perp\ll\kappa_\parallel}$
and a constant, isotropic density $\rho$, 
the conductive term in Eq.~\eqref{eq_dTdt} can be written on the form
\begin{equation}
    \left(\pdv{T}{t}\right)_\mathrm{cond} 
    = \grad_\parallel\cdot 
    \left( \frac{\kappa_{\parallel}^{\ast} }{c_v \rho} T^{5/2} \grad_\parallel T \right) .
    \label{eq_dTdt_cond}
\end{equation}




\section{Self-similar solutions}
\label{app_selfsim}

Here we present a thorough derivation of the self-similar solutions to \change{non}linear diffusion, first described in the seminal paper by \citet{Pattle1959}, given in this paper as Eqs.~(\ref{eq_T}-\ref{eq_chi}). The derivation is based on a 2D-derivation by F.~Moreno-Insertis (2024, priv. comm.) similar to that published in \citet{moreno2022}.

Make the ansatz that the distribution $T$ has the shape and boundary conditions 
\begin{align}
    T(r,t) &= \frac{a_0^m}{a^m}f(\xi)
     , \quad \text{with}~\xi \equiv \frac{r}{a(t)},
     \label{eq_app_Tansatz}\\
    \left.\pdv{T}{r}\right|(r=0,t) &= 0,
    \label{eq_ass_dtdr0} \\
    T[r>R(t);t] &=0.  
    \label{eq_ass_Tinf0}
\end{align}
where $m$ is a constant scale factor, ${a\equiv a(t)}$ is a time-dependent scaling function and ${a_0\equiv a(t=0)}$.
Equation~\eqref{eq_ass_Tinf0} says that the distribution $T$ is zero beyond some finite radius, giving it compact support.
We can choose $m$ so that the solution of Eq.~\eqref{eq_dTdt_intro} has a constant volume integral in $s$ dimensions. The integral out to a radius ${r_\lambda=\lambda a(t)}$, where $\lambda$ is an arbitrary constant, is
\begin{equation}
    \int\limits_{V(r\leq\lambda a(t)} T(r,t) \dd^s r 
    = \Omega_s \!\! \int\limits_0^{\lambda a(t)} \! 
    T(r,t)r^{s-1}\dd r
    = \frac{\Omega_s a_0^m}{a^{m-s}} \!\int\limits_0^\lambda\! f(\xi)\xi^{s-1}\dd\xi,
\end{equation}
where $\Omega_s=\{2,2\pi,4\pi\}$ is the total solid angle for $s=\{1,2,3\}$ dimensions. 
The integral is only constant in time if ${m\equiv s}$.

Next, insert Eq.~\eqref{eq_app_Tansatz} with ${m=s}$ into Eq.~\eqref{eq_dTdt_intro}. By differentiating with respect to (wrt) time and space, followed by isolating all terms with explicit time dependence on the left-hand side (LHS), we end up with 
\begin{equation}
    a^{s \, n+1}\pdv{a}{t} = 
    \frac{-K a_0^{s \, n}
        \left(\xi^{s-1} f^n f'\right)'}{
        \left(\xi^s f\right)'
        },
        \label{eq_app_aadot}
\end{equation}
where the apostrophe marks a partial differentiation with respect to $\xi$, \change{that is} ${f'\equiv\partial f/\partial \xi}$. 
As a consequence of the separation of variables, the RHS is independent of time, making the LHS constant in time. That is achieved when 
\begin{equation}
    a(t) = a_0 \left(1+\chi t\right)^{\tfrac{1}{s \, n+2}},
    \label{eq_app_aoft}
\end{equation}
where $\chi$ is a constant to be determined later, proportional to the RHS of Eq.~\eqref{eq_app_aadot}.

We insert Eq.~\eqref{eq_app_aoft} into Eq.~\eqref{eq_app_aadot} to get
\begin{equation}
    \frac{\chi a_0^2}{s \, n+2} \left(\xi^s f\right)' 
    = -K\left(\change{\xi^{s-1}}f^n f'\right)',
\end{equation}
which we integrate wrt $\xi$ to get
\begin{equation}
    \frac{\chi a_0^2}{s \, n+2} \xi^s f 
    = -K \xi^{s-1} f^n f' + C_1, 
\end{equation}
where the integration constant ${C_1=0}$, because of the boundary condition in Eq.~\eqref{eq_ass_dtdr0}. We make a separation of variables by isolating $f$ on the RHS and integrate once more to get
\begin{equation}
    f^n(\xi) = T_0^n\left(
        1- \xi^2 \frac{n}{s \, n+2}\frac{\chi a_0^2}{2 K T_0^n} 
    \right),
    \label{eq_app_fn}
\end{equation}
where the condition ${f(0)=T(0,0)\equiv T_0}$ has been used to define an integration constant\change{.} 

\change{As} we are interested in a solution that eventually goes to zero, we see that $f(r/a)$ must be zero beyond a radius $R(t)$ defined by setting Eq.~\eqref{eq_app_fn} to zero
\begin{equation}
    R^2(t) =
    \frac{s \, n+2}{n}\frac{2 K T_0^n}{\chi} \left(1+\chi t\right)^{\tfrac{2}{s \, n+2}}
    \equiv R_0^2 \left(1+\chi t\right)^{\tfrac{2}{s \, n+2}},
    \label{eq_app_R2}
\end{equation}
which increases with time for ${\chi>0}$, as expected for diffusion.
Combining the expressions for $a(t)/a_0$ and $f(r/a)$, we get
\begin{equation}
    T(r,t) = 
    T_0 \left(1+\chi t\right)^{-\tfrac{s}{s \, n+2}} 
    \left( 1- \frac{r^2}{R^2}\right) ^{\tfrac{1}{n}}, \text{if $r<R(t)$.}
    \label{eq_app_T} 
\end{equation}
The constant inverse time scale is defined through \eqref{eq_app_R2} as
\begin{equation}
    \chi = \frac{s \, n+2}{n}\frac{2 K T_0^n}{R_0^2}.
\end{equation}
This concludes the derivation of Eqs.~(\ref{eq_T}-\ref{eq_chi}) in Sec.~\ref{sec_theory_ideal}.




\end{appendix}
\end{document}